\documentclass[12pt,preprint]{article}
\usepackage{epsf}
\newcommand{\eqb}{\begin{equation}}
\newcommand{\eqe}{\end{equation}}
\newcommand{\dmb}{\begin{displaymath}}
\newcommand{\dme}{\end{displaymath}}
\newcommand{\pd}{\partial}
\newcommand{\ep}{\varepsilon}
\newcommand{\eab}{\begin{eqnarray}}
\newcommand{\eae}{\end{eqnarray}}

\newcommand{\e}{\mbox{e}}
\newcommand{\be}{\begin{equation}}
\newcommand{\ee}{\end{equation}}

\begin{document}
\begin{titlepage}
\begin{flushright}
TPI-MINN-00/30-T  \\
\end{flushright}
\vspace{0.6cm}

\begin{center}
\Large{{\bf BPS Saturated Vacua Interpolation along One Compact Dimension}}

\vspace{1cm}

R. Hofmann

\end{center}
\vspace{0.3cm}

\begin{center}
{\em Theoretical Physics Institute, Univ. of Minnesota, 
Minneapolis, MN 55455}
\end{center}
\vspace{0.5cm}

\begin{abstract}

A class of generalized Wess-Zumino models with distinct vacua is investigated. 
These models allow for BPS saturated vacua interpolation along one 
compact spatial dimension. 
The properties of these interpolations are studied. 
 
\end{abstract} 

\end{titlepage}

\section{Introduction}

Recently, there has been a great interest in space times with 
$D$ infinite and $n$ compact spatial dimensions 
\cite{Dim1,Dim2}. Phenomenologically, 
this interest originates from the 
idea that the four dimensional Planck mass $M_4$ may be obtained from 
a $4+n$ dimensional fundamental mass $M_{4+n}$ being of the 
order of the electroweak scale, where the $n$ extra spatial dimensions 
are compact. Taking this assertion seriously, the radius $R$ of the extra dimensions 
must be large ($R\le 1$ mm for $n=2$) \cite{Dim1}. In this scenario the (light) 
fields of the Standard Model are trapped on a defect 
along the compact dimensions while gravitons propagate freely. At the scale of distances set 
by the size of the extra dimensions the dynamics underlying the formation 
of the defect(s) and accompanying gravity 
should effectively be describable in terms of field theory 
rather than string theory, although a string realization of low scale gravity 
has been put forward in Ref.\,\cite{Anton}. In a field theoretic context 
defects can appear in the 
form of domain walls, gauge vortices, etc. to have widths 
of the order of the ${4+n}$ dimensional Planck length \cite{Dim1}. 
 In processes with 
 high transverse momentum (comparable to the electroweak scale) 
 it should be possible to lose energy to the bulk region and hencefore violate 
four-momentum conservation in our four dimensional universe. 
A similar picture has been set up for the case of 
infinite extra dimensions in Ref.\,\cite{Dva}. 
There it was argued that the formation of defects in 
supersymmetric theories leading to the spontaneous 
breakdown of a part of SUSY may provide a means of 
trapping of the zero modes and thereby 
reducing the dimension of 
space time for these light modes. 
In contrast to the case of non-compact space time 
the emergence of parallel worlds is conceivable in 
the extra compact dimensions since there 
may be stable constellations of manifold 
defects as we will make explicit below. 

To solidify the above scenario it is important to 
understand the dynamical aspects of defect generation and 
particle trapping. It is also necessary to explain in a detailed way 
how the compact dimensions can be stabilized against a possible 
gravitational collaps or an expansion caused 
by other forces \cite{Sundrum,Dim3}. 

In the framework of supersymmetric theories 
it is known for a long time that the BPS nature of a vacua interpolation preserves a 
part of SUSY and that the very existence of it requires the 
central charge(s) of the SUSY algebra 
to be non-zero \cite{WittenOlive,Gorsky}. Extensive investigations 
of BPS saturated domain walls and their junctions have been performed for example in 
Refs.\,\cite{Brito1,Brito2}. 
For the case of BPS vacua interpolation along one spatial dimension 
the (2+1) dimensional dynamics of light modes due to Nambu-Goldstone fields 
bound to the corresponding domain walls was shown to 
be supersymmetric \cite{Chib}. 

In this paper we revisit the issue of dynamical supersymmetry breaking 
in one compact spatial dimension. The analysis is 
performed for a class of generalized Wess-Zumino toy models in four dimensions 
with one spatial dimension compactified on a circle of radius
$R$. These models admit BPS saturated vacua 
interpolation along the compact dimension. 

The paper is organized as follows: In Section 2 we briefly review the 
pecularities of BPS saturation for solutions dependent on one compact, spatial dimension 
as they were worked out in Ref.\,\cite{Losev}. We then introduce a class of 
generalized Wess-Zumino models with members labeled by an integer $N$. 
The $N$th model has $N+1$ vacua which can be connected by BPS 
saturated solutions of the scalar sector. 
The properties of these vacua interpolations 
are investigated on the classical level, 
and some remarks 
about quantum corrections are made. Section 3 summarizes the results and comments 
on how the toy models considered here relate 
to the phenomenological notion of 
large extra dimensions.

\section{Model and BPS solutions}

In the case of (3+1)-dimensional non-compact 
space the vacua interpolating solutions 
of a supersymmetric model with a set of distinct ground states 
are topologically stable \cite{Losev}. These solutions may or 
may not be BPS saturated. Specializing to a (3+1)-dimensional 
Wess-Zumino model with superpotential $\cal W$, 
one may seek for an interpolation of the vacua $\Phi=\Phi(z)$ 
\footnote{In the following we will use the same symbol for the 
chiral superfield and the scalar component field.} satisfying 
the following equations of BPS saturation \cite{Chib} 
\eqb
\label{BPS}
\pd_{z}\Phi(z)=\frac{\pd}{\pd\bar\Phi}\,{\bar{\cal W}}\ ,\ \ 
\pd_{z}{\bar\Phi}(z)=\frac{\pd}{\pd\Phi}\,{\cal W}\ ,
\eqe
where a possible phase factor has been absorbed 
in the definition of the superpotential. 
If such a solution exists then its tension $\tau[\Phi]$ is equal to the 
$(1,0)$ central charge $Z$ of the 
SUSY algebra, which has been shown long ago \cite{WittenOlive}.

Since $\tau[\Phi]$ is given by the 
integral over a smooth, positive semi-definite function 
the nonvanishing of $Z$ is a necessary condition for the existence of 
a BPS saturated vacua interpolation. It is given as \cite{Chib} 
\eqb
Z=\int dz\, \frac{d}{dz}\, {\bar{\cal W}}(\Phi)\ ,
\eqe
where the integration extends over the entire dimension.  

In the case of a (3+1) dimensional space with the $z$ coordinate compactified 
on a circle with radius $R$ the field $\Phi$ has to 
satisfy periodic boundary conditions $\Phi(0)=\Phi(L\equiv 2\pi R)$. Hence, 
the vanishing of $Z$ can only be avoided if the 
superpotential ${\cal W}$ is a multivalued function, 
i.e. the target manifold of field 
variables admits non-contractable cycles. 
The former possibility is unproblematic as long as the derivatives  
${\cal W}^\prime\equiv\frac{\pd{\cal W}}{\pd \Phi},\ {\cal W}^{\prime\prime}
\equiv\frac{\pd^2{\cal W}}{\pd \Phi^2},\ \dots$ are single 
valued since it is these quantities that 
define the physically relevant 
scalar potential $V(\Phi,{\bar \Phi})\equiv \frac{\pd{\cal W}}{\pd \Phi}\frac{\pd{\bar{\cal W}}}{\pd\bar\Phi}$ 
and the fermionic interactions \cite{Chib}. 
 
Let us consider the one-parameter set of generalized Wess-Zumino 
models with the standard K\"ahler and the following superpotential 
\eqb
\label{model}
{\cal W}\equiv i\left(\log \Phi-\frac{1}{N+1}\, \Phi^{N+1}\right)\ ,\ \ \ \ (N=1,2,\dots)\ .
\eqe
The vacua $\Phi^*_{N,k}$ of model $N$ are determined by the zeros of 
${\cal W}^\prime$, and hence we have 
$\Phi^*_{N,k}=\e^{\frac{2\pi i}{N+1}k}\ ,\ (k=0,\dots,N)$. For 
$N=1,2,3$ the scalar potential 
$V(\Phi,{\bar\Phi})={\tilde V}(\mbox{Re}\Phi,\mbox{Im}\Phi)$ is depicted in Fig. 1.
\begin{figure}
\vspace{7cm}
\includegraphics{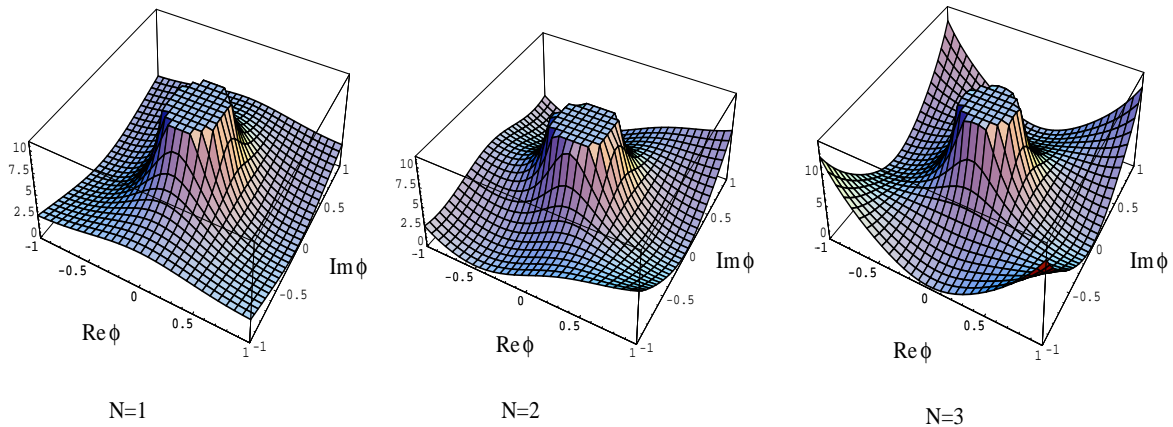}
\caption{The scalar potential for the cases $N=1,2,3$.} 
\label{} 
\end{figure}
We are interested in BPS saturated vacua interpolation $\Phi=\Phi(z)$. 
Using the BPS-equations (\ref{BPS}), 
it is easy to see that the quantity 
\eqb
\label{I}
I\equiv-\mbox{Im}\,{\cal W}=-\log|\Phi|+1/2\left[(\mbox{Re}\,\Phi)^2-(\mbox{Im}\,\Phi)^2\right]\ ,\ \ 
(N=1)\ ,
\eqe
is a constant of "motion" \cite{Chib,Losev}. From Eq.\,(\ref{I}) it is clear that 
for initial values $|\Phi(0)|\ll 1$ the BPS 
trajectories are circles centered at the origin. Indeed, in 
this limit the pole in ${\cal W}^\prime,\ \bar{\cal W}^\prime$ 
is dominant (right-hand sides of (\ref{BPS})), and 
the BPS equations have the 
simple winding solution \cite{Losev,SD}
\eqb
\label{wi}
\Phi_{wi}=\sqrt{L/2\pi}\,\e^{-\frac{2\pi i}{L} z}\ ,
\eqe
which implies the relation $L=2\pi A^2$ between the 
initial value $A\equiv\Phi_{wi}(0)$ ($A$ real) 
and the cycle length $L$. There are similar 
solutions with winding number $n>1$. All these 
solutions are uninteresting for the purpose of the present paper 
in the sense 
that (a) they do not interpolate between the vacua of the models, 
and (b) they have a uniform distribution of
energy density.

In the case $A=1-\delta$ ($0<\delta\ll 1$) 
Fig. 2 depicts the trajectories of the 
solutions to Eq.\,(\ref{BPS}) for $N=1,2,3$. 
\begin{figure}
\vspace{7cm}
\includegraphics{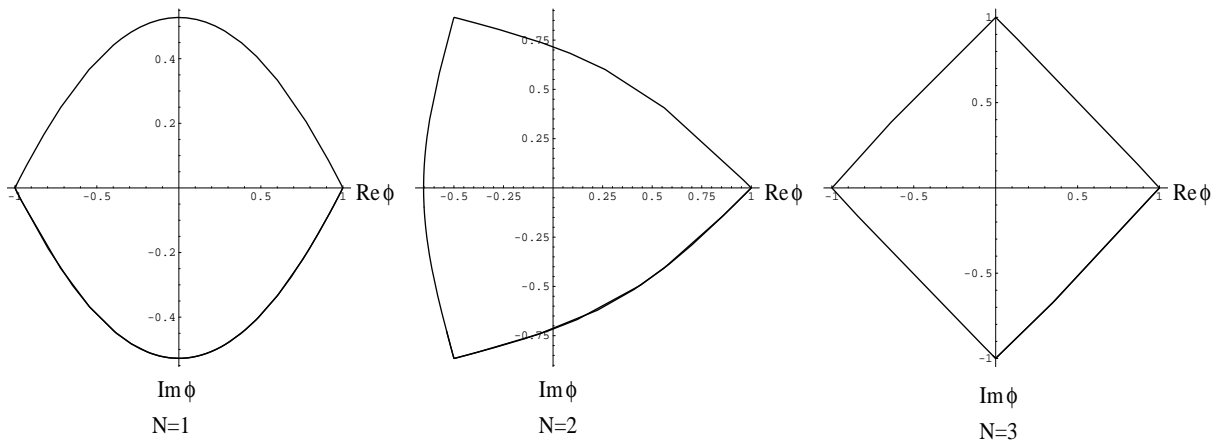}
\caption{The BPS vacua interpolations for $N=1,2,3$.} 
\label{} 
\end{figure}
Winding around the pole at $\Phi=0$, these solutions (approximately) 
connect all vacua of the corresponding model 
in a clockwise sense. We argue later that for each $N$ there is a bijection between the 
BPS saturated periodic solutions of a given topology $n$ and the cycle length $L$, 
which hence can be used to label these solutions. In the periodic case 
the tension $\tau[\Phi_L]$ is the same for all $N$ and $L$. 
It is given by minus the residue of the pole of 
${\cal W}^\prime$ times $2\pi i$, that is $\tau[\Phi_L]\equiv 2\pi$. 
Fig. 3 shows Re$\,\Phi_L$ and the 
energy density $\ep_L$ as functions of $z$.  
\begin{figure}
\vspace{8.8cm}
\includegraphics{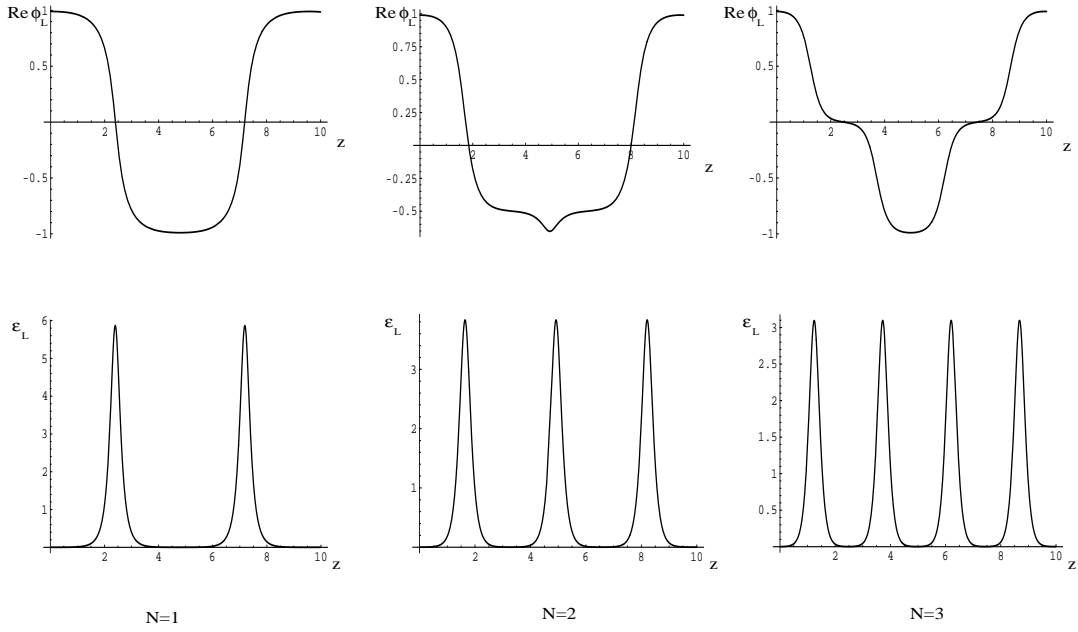}
\caption{The real part and the energy density $\ep_L$ of the vacuum interpolating 
BPS solutions $\Phi_L$ for $N=1,2,3$.} 
\label{} 
\end{figure}
In order to quantify the behavior of the solutions in the 
vicinity of the vacua $\Phi^*_{N,k}$ ($k=0,\dots,N$) for the topological sector $n=1$ 
we may linearize the BPS equations 
about these. Linearizing about $\Phi^*_{N,0}=1$ and demanding the initial condition 
$\tilde{\Phi}(0)=(A-1)$ for the deviation $\tilde{\Phi}$, 
we obtain the following solution 
\eqb
\label{lin}
\tilde{\Phi}(z)=(A-1)\,\left[\cosh\left((N+1)z\right)+i\sinh\left((N+1)z\right)\right]\ .
\eqe
Approximating $\Phi(z)$ at $z=\frac{L}{2(N+1)}$ (first wall) as
\eqb
\label{ap}
1+\tilde{\Phi}\left(\frac{L}{2(N+1)}\right)\approx
\Phi\left(\frac{L}{2(N+1)}\right)\approx \cos(\pi/(N+1))\ \e^{-\frac{\pi i}{N+1}}\ ,\ \ (N>1)\ , 
\eqe
and, by means of Eq.\,(\ref{lin}), comparing real parts we obtain 
in the limit of large $L$ ($A\to 1$) the 
relation between initial value $A$ and cycle length $L$  
\eqb
\label{A-L}
L=2\, \log\left[\frac{1-\cos^2(\pi/(N+1))}{1-A}\right]
\approx\, \log\left[\frac{\pi^4}{N^4(1-A)^2}\right]\ ,\ \ (N\gg 1,\,N^4\ll 
\frac{\pi^4}{(1-A)^2})\ . 
\eqe
For small $N$ Eq.\,(\ref{A-L}) indicates a weak dependence 
of $L$ on $N$ in 
accord with the numerical findings of Fig. 3. 
In Fig. 4 we indicate $L$ as a 
function of the initial value $A$ for $N=1$. The quadratic regime of the pole
approximation is present up to values as high as $A=0.8$. 
The above numerical calculation yields a stronger 
increase of $L$ than does the pole approximation. 
The logarithmic regime starts at $A\approx 0.999$.  
\begin{figure}
\vspace{7cm}
\includegraphics{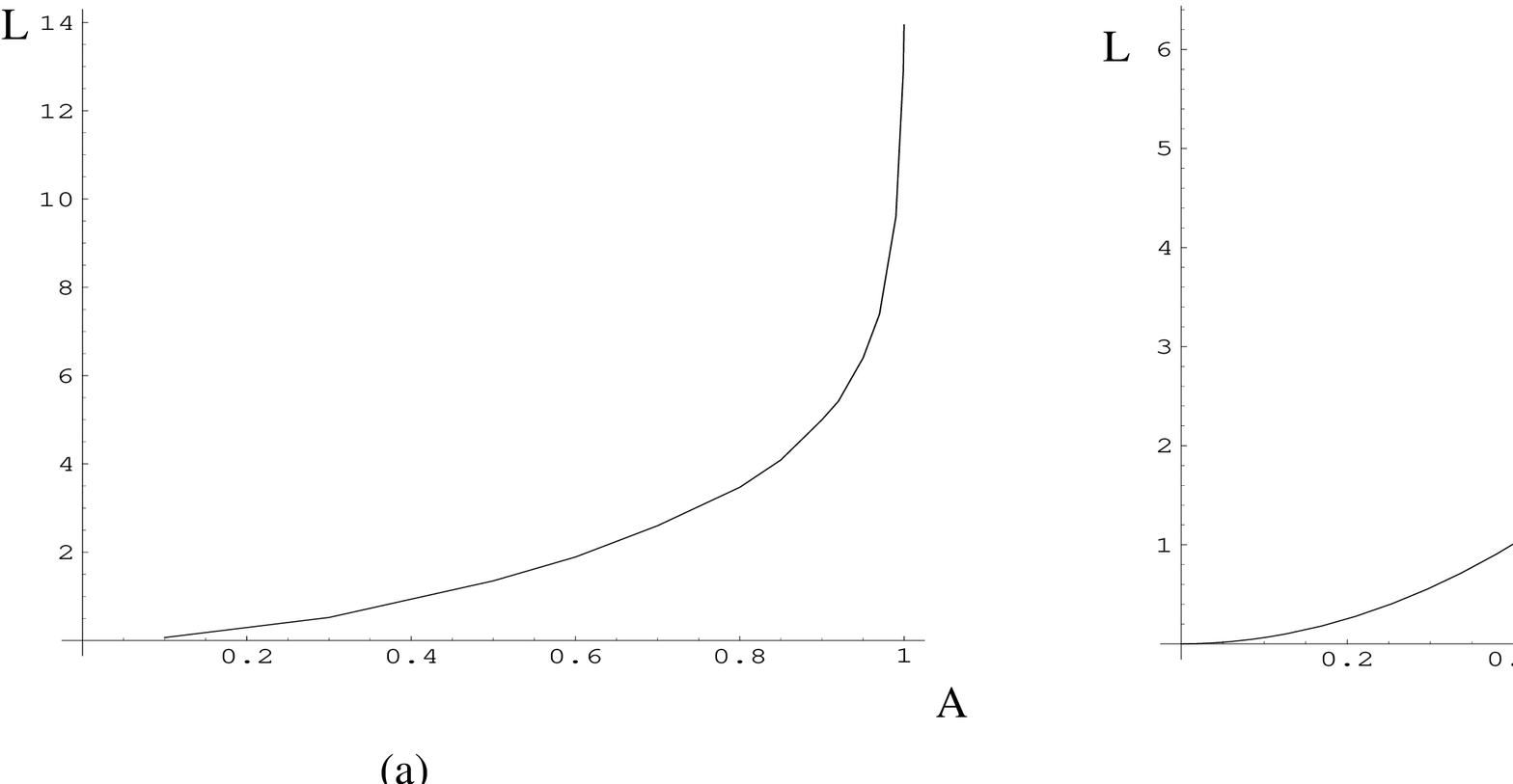}
\caption{The cycle length $L$ as a function of the initial value $A$ for $N=1$. 
The case (a) is the numerical result whereas 
(b) corresponds to the winding solutions of the pole approximation ($L=2\pi A^2$).} 
\label{} 
\end{figure}

In the large $N$ limit (with the condition $N\gg 1,\,N^4\ll \pi^4/(1-A)^2$ satisfied) 
the linearization of the BPS 
equations is justified for an entire 
interpolation of two adjacent vacua. Then, from Eqs.\,(\ref{lin}) and 
(\ref{A-L}) we conclude that 
the energy density $\ep^{w}_L$ in the center of the wall is 
\eqb
\ep^{w}_L\approx \frac{\pi^4}{N^2} \ ,\ \ \ (N\gg 1,\,N^4\ll \pi^4/(1-A)^2)\ ,
\eqe
that is, $\ep^{w}_L$ is independent of the initial value $A$. 
We know that the tension $\tau$ is independent of $N$ and $L$. 
Defining the width $\Gamma$ of each wall by
\eqb
\tau=(N+1)\times\ep^{w}_L\times \Gamma\ ,
\eqe
we find
\eqb
\Gamma\approx \frac{2}{\pi^3}N\ ,\ \ \ (N\gg 1,\,N^4\ll \pi^4/(1-A)^2)\ . 
\eqe
Hence, in the above limit the sequence of 
domain walls looses its peak structure 
suggesting that there is a critical ratio $R=\pi^4/((1-A)^2 N^4)\gg 1$ 
above which the individual walls 
melt into an almost uniform energy distribution along the 
entire compact dimension.  

What happens for $A>1$? The solution of the linearized 
problem of Eq.\,(\ref{lin}) indicates that 
$\Phi(z)$ moves away from the origin within the first quadrant. 
For $N=1$ let us assume 
that at some $z_0>0$ $\Phi$ is close to 
the line $\e^{i\pi/4}t\ ,\ (0\le t<\infty)$, that is
\dmb
\Phi(z)\approx f(z)(1+i)\ ,\ \ (f\ \mbox{positiv real for}\ z\ge z_0)\ .
\dme
Since $\Phi(z)$ initially moves away 
from the origin we furthermore 
can assume that $f$ is considerably larger than unity. 
Omitting then the pole term in the right-hand side of the BPS
equations, we deduce that $\Phi(z)=f(z)(1+i)$ is an approximate 
solution for $z\ge z_0$ with $f(z)=\e^{z}$. 
This blow-up behavior is observed numerically also for $N=2,3$. 
The corresponding solutions are not periodic, and therefore 
they are not relevant to the case of a compactified dimension. 
For $N=1$ (and presumably also for $N>1$) we thus have found 
that the cycle length $L$ of the periodic (non-constant) BPS solutions as a function of the 
initial value $A$ is a bijection $L:\,(0,1)\rightarrow\,(0,\infty)$. 
The set of periodic BPS solutions at given $L$ is 
therefore uniquely indexed by the winding number $n=1,2,\dots$.
\footnote{There is also a bijection between the periodic
BPS solutions and the initial value $A\in(0,1)$: Since the right-hand 
sides of the BPS equations (\ref{BPS}) are smooth in the entire complex plane except for the point zero 
the solutions are uniquely determined by the initial value. It remains to be shown that 
{\em each} periodic solution is associated with one value of $A\in(0,1)$. 
Assume this was not the case, 
that is, there is a periodic solution $\Phi_B(z)$ with 
initial value $B\notin\{\mbox{trajectories labeled by}\ A\}$. 
At some $z_R>0$ $\Phi_B$ must cross the real axis with 
$0<\Phi_B(z_R)<1$ for otherwise it will be not periodic as we showed above. 
Setting $A\equiv\Phi_B(z_R)$, we obtain a contradiction to the assumption
that $B$ coincides with no point of the trajectories labeled by $A$.} 
Rescaling the K\"ahler metric, we therefore conclude
that BPS saturation is possible in all regimes of coupling strength (see Ref.\,\cite{Losev} for details).      
We can view the BPS interpolation $\Phi_L(z)$ at cycle length 
$L$ as the classical ground state of the model. It is known for a long time 
that the tension of this configuration does not receive any 
perturbative corrections \cite{Chib}. Since the mass of the BPS interpolation is infinite, nonperturbative 
effects like instanton-tunneling can not change the BPS tension either. 

Regarding the superpotential of 
Eq.\,(\ref{model}) to define a 
generalized Landau-Ginzburg model \cite{Losev} with 
extended ${\cal N}=2$ SUSY in (1+1) dimensions (spatial dimension compact) \cite{Losev},
 the infinite-mass argument would no longer hold. A priori, there could be instanton mediated tunneling 
 to change the energy of the state and make it non-BPS \cite{Losev,Witten}. 
However, since we have a bijection 
between cycle length and periodic solution there is 
only one BPS state with given winding number $n$ 
at a given radius $R$. Hence, non-perturbative 
quantum tunneling within one topological 
sector is impossible.

\section{Summary}

The existence of BPS saturated vacua interpolations along one 
compact spatial dimension has been exemplarily demonstrated 
for a class of generalized Wess-Zumino models in (3+1) dimensions. 
The non-vanishing of the $(1,0)$ central charge $Z$ of 
the SUSY algebra, necessary for the existence of 
BPS saturated solutions, is achieved by a 
superpotential with a branch cut. The solutions of the BPS equations 
are either periodic (tension=$2\pi$) or blow-up (infinite tension). 
For the case $N=1$ we showed explicitely that 
there is a bijection between the set of periodic 
solutions and the cycle length $0<L<\infty$. In the limit of 
large $L$ and $N$ the width $\Gamma$ of the walls 
scales as $N$. On the other hand, for fixed initial 
value $0<A<1$ the length of the cycle scales logarithmically with $N$ suggesting 
that there are critical combinations of the 
quantities $N$ and $A$ for which the individual walls start to 
melt into a uniform distribution of energy 
throughout the entire compact dimension. 

Viewing the above class of 
models as ${\cal N}=2$ generalized Landau-Ginzburg 
models in (1+1) dimensions \cite{Losev}, 
the energy of the 
classical BPS state of 
cycle length $L$ does not get affected 
by non-perturbative instanton 
tunneling for $N=1$ (presumably this is also true for $N>1$). 
Perturbative corrections in general 
change the shape of the BPS interpolation 
but leave its energy 
(tension) intact \cite{Chib}. 

The class of $(3+1)$ dimensional models that were investigated in this 
paper are toy models 
in the following sense: Since four is 
the maximal dimension at which one can build a non-trivial, 
supersymmetric theory of scalars and fermions Wess-Zumino models cannot 
make direct contact with the phenomenology of extra dimensions 
as put forward in Refs.\,\cite{Dim1,Dim2,Anton}. However, they do provide 
a simple framework for the visualization of compactification scenarios. 
The possibility of the dynamical generation of a stable 
"brane-lattice", which may cause a gravitational stabilization of 
the extra dimension \cite{Dim3}, is transparent in the 
setting of a Wess-Zumino model. 
Concerning the question of SUSY breaking in the universe 
of light particles trapped on a spontaneously generated 
defect, one may introduce small SUSY 
breaking terms into the scalar potential, 
which allow for a perturbative treatment and periodicity of 
the perturbed solution. Alternatively, one 
could start with a non-supersymmetric model 
allowing for BPS saturated 
domain walls. The reason for the existence of 
light, localized fermions is 
then solely due to generalized 
index theorems \cite{JR} in 
contrast to the case of a spontaneous, partial 
breakdown of SUSY, where the localized 
fermions primarily are Goldstinos \cite{Chib}.

\section*{Acknowledgements}    

The author likes to thank M. Shifman for 
reading the manuscript and suggesting 
various improvements. Stimulating and useful 
conversations with T. ter 
Veldhuis are gratefully acknowledged. This work was funded 
by a postdoctoral fellowship of 
Deutscher Akademischer Austauschdienst (DAAD).

\bibliographystyle{prsty}

\end{document}